\documentclass{ws-ijmpb}
\usepackage{multicol}
\usepackage{graphicx}
\usepackage{float} 
\usepackage{comment}
\usepackage{xcolor}
\begin{document}

%
\catchline{}{}{}{}{}
%

\title{Effective quantum tunneling from a semiclassical momentous approach}

\author{L. Arag\'on-Mu\~{n}oz}

\address{Instituto de Ciencias Nucleares, Universidad Nacional Aut\'onoma de M\'exico\\
Ciudad Universitaria, 04510 Coyoac\'an, Ciudad de M\'exico, M\'exico\\
luis.aragon@correo.nucleares.unam.mx}

\author{G. Chac\'on-Acosta}

\address{Departmento de Matem\'aticas Aplicadas y Sistemas, Universidad Aut\'onoma
	Metropolitana-Cuajimalpa\\
	 Av. Vasco de Quiroga 4871, Ciudad de M\'exico, 05348,
	M\'exico\\
	gchacon@cua.uam.mx}

\author{H. Hernandez-Hernandez}

\address{Facultad de Ingenier\'ia, Nuevo Campus Universitario, Universidad Aut\'onoma de Chihuahua\\
	Chihuahua 31125, M\'exico\\
	hhernandez@uach.mx}

\maketitle

\begin{history}
\received{Day Month Year}
\revised{Day Month Year}
\end{history}

\begin{abstract}
In this work, we study the quantum tunnel effect through a potential barrier within a semiclassical formulation of quantum mechanics based on expectation values of configuration variables and quantum dispersions as dynamical variables. The evolution of the system is given in terms of a dynamical system for which we are able to determine numerical effective trajectories for individual particles, similar to the Bohmian description of quantum mechanics. We obtain a complete description of the possible trajectories of the system, finding semiclassical reflected, tunneled and confined paths due to the appearance of an effective time-dependent potential. 
\end{abstract}

\section{Introduction}

Quantum tunneling can be considered as one of the defining phenomena of quantum mechanics because it is a direct consequence of its probabilistic wave structure. It occurs when a particle can penetrate a potential barrier with a non-zero probability, even when the maximum of the potential is larger than the kinetic energy of the particle, thus this microscopic phenomenon has no counterpart in classical physics. Since the beginning the tunnel effect was proposed as a physical mechanism that explains both the field electron emission\cite{electron_emission} and the alpha decay\cite{alpha-decay}, which were the first successful applications of quantum theory to nuclear phenomena\cite{Bethe}. This was followed by many other important applications:  analysis of the spectra of molecules\cite{moleculas}, molecular and condensed matter physics\cite{condensed}, the tunnel diode and electron tunneling in solids\cite{tunel-solidos}, superconductivity engineering, and manufacturing of aggregates on the nanoscopic and mesoscopic scale\cite{superconductivity}, etc. All this has resulted in a fruitful exchange between experimental techniques and theoretical concepts, in physics and chemistry of atomic, molecular, and condensed matter\cite{libros-qm-apps}, among many others. In all these systems it is necessary to study the tunnel effect in interaction with some potentials and to solve the Schr\"odinger equation subject to boundary conditions, which in general is a quite complicated matter. However, in certain regimes, it is possible to implement semi-classical approximations where the solutions are given as perturbations to known classical solutions. This, and the exploiting of the intuition acquired from classical physics, serves to implement numerical methods.

In this work we analyze the quantum tunneling effect using a semiclassical approach to quantum mechanics\cite{Bojo1} which resembles the description given by Bohm\cite{Bohm}, in which the evolution of the system is given by an ensemble of trajectories. An interesting feature of the Bohmian description is that one can assume some initial conditions for trajectories as in any dynamical system and then one determines its evolution and its characteristics, such as the tunneling time. Since Bohm mechanics is based on an ensemble of trajectories the system is subject to statistical fluctuations. Here we use an alternative approach based on one-particle trajectories on an enlarged phase space consisting of the expected values of quantum observables and of all higher-order quantum dispersions. A similar method has been used in quantum chemistry where ionization and chemical reactions are studied\cite{Prezhdo}. The expected values of quantum variables in a given state can be directly interpreted as classical canonical variables, while dispersions form a new set of dynamical variables that can be considered as statistical moments. Their dynamics is governed by a highly coupled dynamical system. For the tunneling problem the interaction of both classical variables and quantum dispersions effectively modifies the classical potential barrier yielding a quantum effective potential. In this way states that do not appear in the classical setting may appear allowing particles to be reflected, tunnel the barrier, or get trapped.

In this effective model one can study which of the individual trajectories cross the barrier, even if at a later time it crosses back or gets trapped, as we will show. Although quantum trajectories are not experimentally observable valuable qualitative information can still be extracted from this approach, that could be used, for instance, in the design of quantum control experiments involving this effect \cite{tunnel_control}. This model has also proved its effectiveness in the semi-classical description of some quantum cosmological models\cite{Bojowald2}, provides a very interesting tool to analyze the concept of tunneling times\cite{time-tunnel,Bojo-Crowe}, and can be applied to more general scenarios\cite{quantum_kepler}.

The paper is organized as follows: In section \ref{method} we describe the effective momentous description of quantum mechanics, that is, the semi-classical description of a quantum system as a Hamiltonian dynamical system of expected values with infinitely many degrees of freedom. In section \ref{classical}, we describe the classical potential barrier for which we will study the tunnel effect, whereas, in section \ref{trajectories}, the initial conditions for the evolution of the system are given. In section \ref{numerics} we numerically solve the truncated system up to the second-order  (the third-order is presented in the Appendix),  and we describe the behavior of the system in terms of effective trajectories and the interpretation of transmission and reflection. Finally, in section  \ref{discussion} we discuss our results and propose some future extensions.

\section{Effective dynamics of quantum systems: Momentous quantum mechanics} \label{method}

It is well known that quantum mechanics has had a successful impact on the development of science and technology, however, because it being a probabilistic theory with intricate dynamics, the study of general systems turns out to be very complicated and the use of semi-classical approximations is a very useful tool. Semiclassical approximations are important to extract new effects or potentially observable phenomena in regimes in which quantum effects become relevant. Several such approximations have been developed\cite{libros-qm-apps,Prezhdo,bouncing-ball}, especially the one known as semiclassical momentous quantum mechanics\cite{Bojo1}.

This formulation is based on a set of semiclassical equations for a dynamical system, which is equivalent to the full quantum description. The equations of motion are obtained using an effective Hamiltonian that considers the expected values of the observables, their associated moments, and the quantum dispersions as classical configuration variables. Quantum fluctuations are introduced as a set of new semiclassical dynamical variables, defined as follows
\begin{equation}\label{1}
G^{a,b} \equiv \left\langle (\hat{x}-x)^a (\hat{p}-p)^b \right\rangle_{\textrm{\tiny{Weyl}}},
\quad a+b \geq 2
\end{equation}
where $x\equiv\langle \hat{x}\rangle $ and $p\equiv\langle
\hat{p}\rangle $, and the subscript indicates that the operators
inside the brackets are Weyl ordered.
In this way we obtain a classical evolution that gets back-reacted by the quantum variables 
(or dispersions) and hence modified.

The quantum dispersions are not completely arbitrary, they are subject to generalized uncertainty
relations such as
\begin{equation}\label{uncertainty}
G^{2,0}G^{0,2} - (G^{1,1})^2 \geq \frac{\hbar^2}{4},
\end{equation}
where $G^{2,0}$ and $G^{0,2}$ are the standard dispersions
$\Delta x^2$ and $\Delta p^2$, and (\ref{uncertainty}) simplifies to the
usual uncertainty principle for pure states\cite{libros-qm}.

Evolution is obtained in the usual way, by evaluating Poisson brackets of a dynamical variable $f$ with the \emph{quantum} effective Hamiltonian $\dot{f} = \{f,H_Q\}$, which is defined as the expectation value of the standard Hamiltonian operator
\begin{equation}\label{Hamiltonian_eff}
\langle \hat{H} \rangle    \equiv H_{Q} = H(x,p) + \sum_{a,b} \frac{1}{a!b!} \frac{\partial^{a+b} H}{\partial x^a \partial
	p^b} G^{a,b},
\end{equation}
where $H(x,p)$ is the classical Hamiltonian. Quantum variables $G^{a,b}$ are now dynamical, as also are the conjugate canonical variables ($x, p$). Evidently, for general systems, a dynamic system of infinite dimensions is obtained which, although complicated, provides a complete description of the system.

As mentioned above, the equations of motion are obtained computing the Poisson brackets of the corresponding variable with the quantum Hamiltonian: for systems with one degree of freedom we obtain
\begin{eqnarray}
\{x,p\} &=& 1, \label{4.1}\\
\{x,G^{a,b}\} &=& 0, \label{4.2}\\
\{p,G^{a,b}\} &=& 0. \label{4.3}
\end{eqnarray}
The Poisson algebra for the moments was obtained in general in \cite{Bojowald2} 
\begin{eqnarray}\label{5}
\{G^{a,b},G^{c,d}\} &=& ad G^{a-1,b} G^{c,d-1} - b c G^{a,b-1}
G^{c-1,d} + \nonumber \\
&+& \sum_{n} \left(\frac{i \hbar}{2}\right)^{n-1}
K^n_{abcd} G^{a+c-n,b+d-n},
\end{eqnarray}
where the sum runs over odd numbers from $n=1 \ldots \tilde{N} $,
with $1 \leq \tilde{N} < \min[a+c,b+d,a+b,c+d]$, and the coefficient
is

\begin{equation}\label{6}
K^n_{abcd} = \sum_{s=0}^n (-1)^s s!(n-s)! \left(
\begin{array}{c}
a \\
s \\
\end{array}
\right) \left(
\begin{array}{c}
b \\
n-s \\
\end{array}
\right) \left(
\begin{array}{c}
c \\
n-s \\
\end{array}
\right)\left(
\begin{array}{c}
d \\
s \\
\end{array}
\right).
\end{equation}

This method has been successfully applied to several interesting physical systems, most notably to the quantum harmonic oscillator, where it was shown that the ground state energy is added to the classical Hamiltonian \cite{Bojo1,LQC-dy}, and also to anharmonic systems where adiabatic approximations were proven useful. It has also been applied to isotropic and homogeneous models in loop quantum cosmology \cite{MB,MBNat}, where one can analyze several cosmological descriptions including matter \cite{MB-H}, cosmological constant, among others.

It is also possible to generalize the above formulae for systems with more than one pair of canonical conjugate variables (see for instance \cite{Bojowald2}), however, for our analysis the description with one degree of freedom will suffice. 


\section{Classical model of the tunnel effect} \label{classical}

\begin{figure}[H]
	\centering
	\includegraphics[scale=0.42]{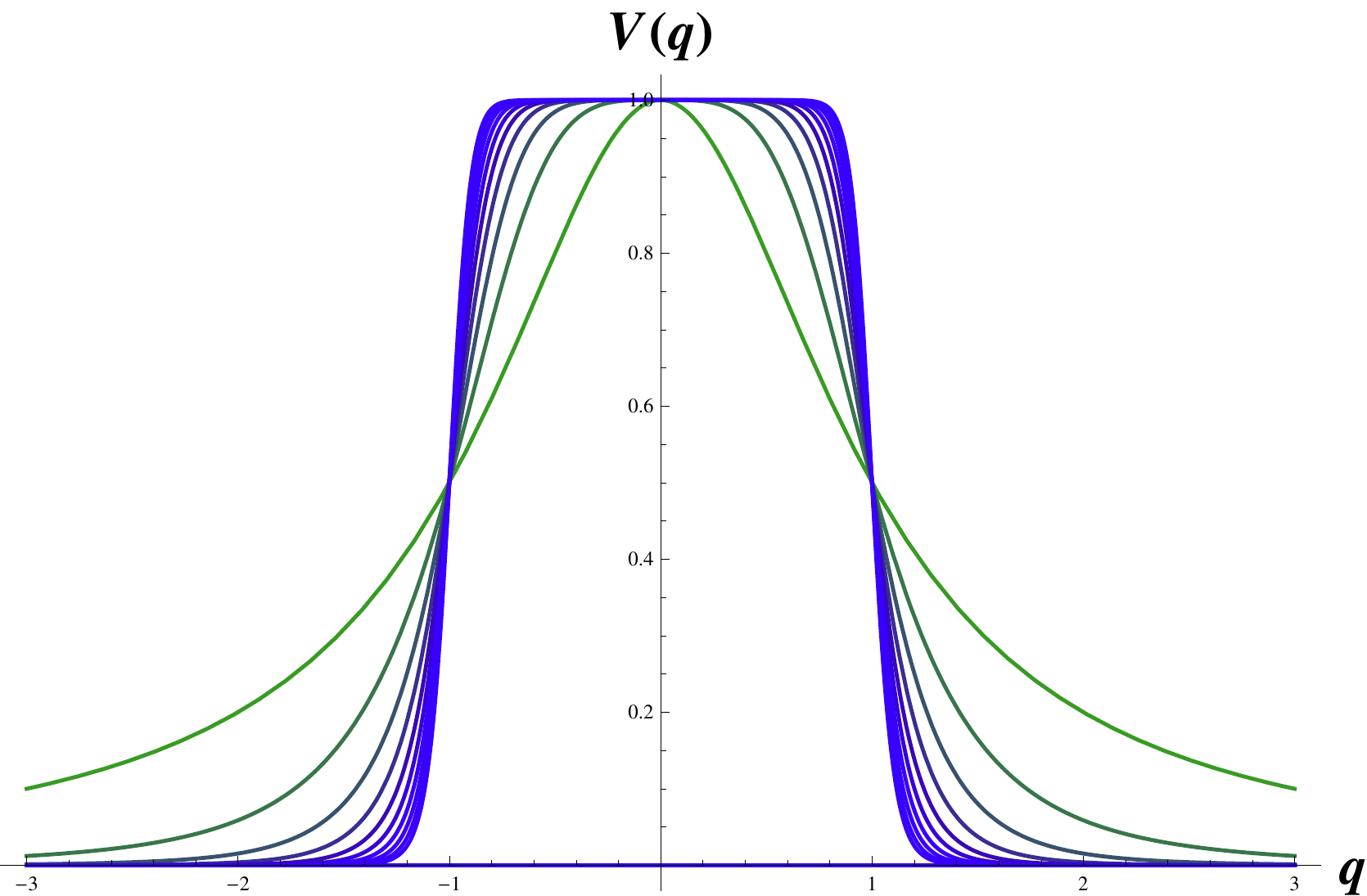}
	\caption{Smoothed potential $V(q)$ for $a\,=\,1$, $\alpha=1$, from $n=1$ (green) to $n=10$ (blue)}
	\label{class_potential}
\end{figure}

In this section we want to analyze the tunnel effect for particles scattering a one dimensional potential barrier, a system with one degree of freedom, so we will describe the classical scenario. This will clarify some aspects that we will use when we study the quantum case. We consider smooth potentials in our effective description (\ref{Hamiltonian_eff}), although this requirement is not required. The system consists of particles of mass $m$ in the presence of a family of potentials of the form
\begin{equation}
\label{potential}
V(q)\,=\,\displaystyle\frac{\alpha}{q^{2\,n}\,+\,a^{2\,n}},\qquad\qquad\qquad \alpha,\,\,a\,\in\,\mathbb{R},\qquad n\,\in\,\mathbb{N}.
\end{equation}
with height $V_0=\alpha/a^{2\,n}$ at $q=0$, and $a$ being the width of the potential. It can bee seen from Fig.\ref{class_potential}
that the potential looks more like the standard potential barrier for large $n$.

Classical dynamics is obtained from Hamilton equations
\begin{eqnarray} 
\dot{q} &=& \{q,H\}\,=\,p/m; \label{eom_q} \\ 
\dot{p} &=& \{p,H\}\,=\,-\,V', \label{eom_p_class}
\end{eqnarray}
with Hamiltonian $H= \frac{p^2}{2m} +V(q)$. Classical return points are given by the condition 
\begin{equation} \label{return_points}
V(q)=\frac{\alpha}{x^{2n}+a^{2n}}=E\; \Rightarrow \quad x=a (\gamma -1)^{1/ 2n},
\end{equation}
where we introduce the parameter $\gamma\,=\,V_{0}/E$. 

Classically there is no tunneling since a particle incoming from the left either will bounce back off the potential for $E< V_0$ or pass over the barrier when $E>V_0$, but could not go ``through" the barrier.

Equation (\ref{eom_q}) gives the following equation of motion 
\begin{equation}
\label{E4}
p\,=\,m\,\dot{q}\,=\,\left[2\,m\,E\,\left(\displaystyle\frac{q^{2\,n}-x^{2\,n}}{q^{2\,n}+a^{2\,n}}\right)\right]^{1/2},\quad |x|\,\leq\,|q|,\quad \text{and }\gamma \geq 1.
\end{equation}
There are no analytic solutions to this equation for arbitrary values of $n$ and $\alpha$ although it is possible to obtain an approximate expression for $q(t)$ in the limiting case when $\gamma=1$ given by
\begin{equation}
\left(\displaystyle\frac{q_{\odot}}{q_{\circ}}\right)^{n-1}\approx \left[\!\!\left(\displaystyle\frac{2V_{0}q_{\circ}^{2(n-1)}}{ma^{2}}\right)^{1/2}\!\!\!\!\!(n-1)\,t\,+\,_{2}F_{1}\left[\begin{array}{r}
\frac{1-n}{2n},-\frac{1}{2};\\
\frac{1+n}{2n},\end{array}\!\!\!\!-q_{\circ}^{4}\right]\right]^{-1}\,\left[1\,+\,\displaystyle\frac{1-n}{2(1+n)}\,q_{\odot}^{2n}\right],
\end{equation}
where $q_{\odot}=|q/a|$, $q_{\circ}=|q_{i}/a|$, $q_i$ are the initial positions of the particle and $\,_{2}F_{1}$ is the hypergeometric function.

\begin{figure}[H]
		\includegraphics[scale=0.6]{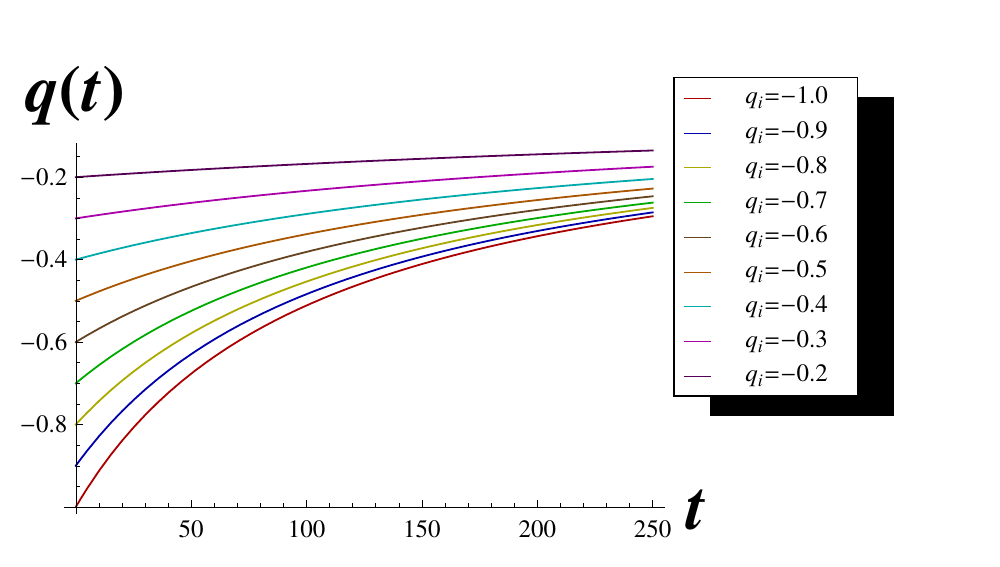}
		\caption{Classical trajectories for $\gamma=1.00$.}
		\label{class_traj_1}
	\end{figure}

\begin{figure}[H]
		\includegraphics[scale=0.6]{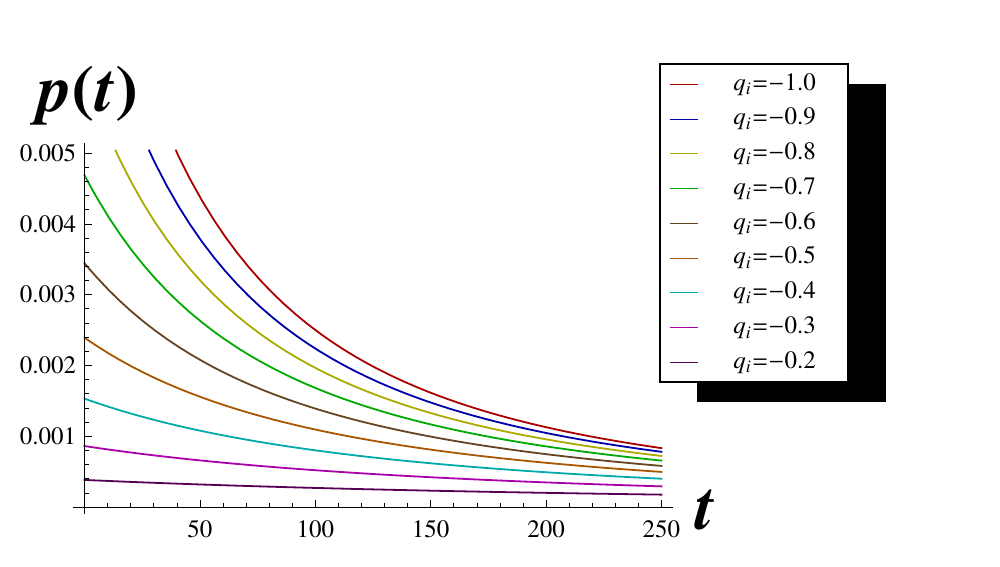}
		\caption{Classical momentum for $\gamma=1.00$.}
		\label{class_mom_1}
	\end{figure}

\begin{figure}[H]
		\includegraphics[scale=0.6]{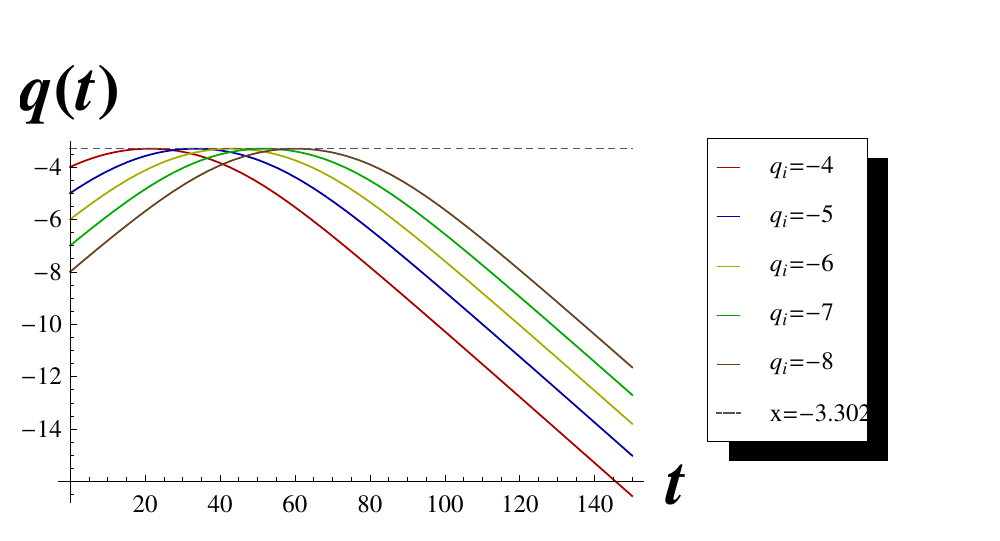}
		\caption{Classical trajectories for $\gamma=1.46484$.}
		\label{class_traj_2}
	\end{figure}

\begin{figure}[H]
		\includegraphics[scale=0.6]{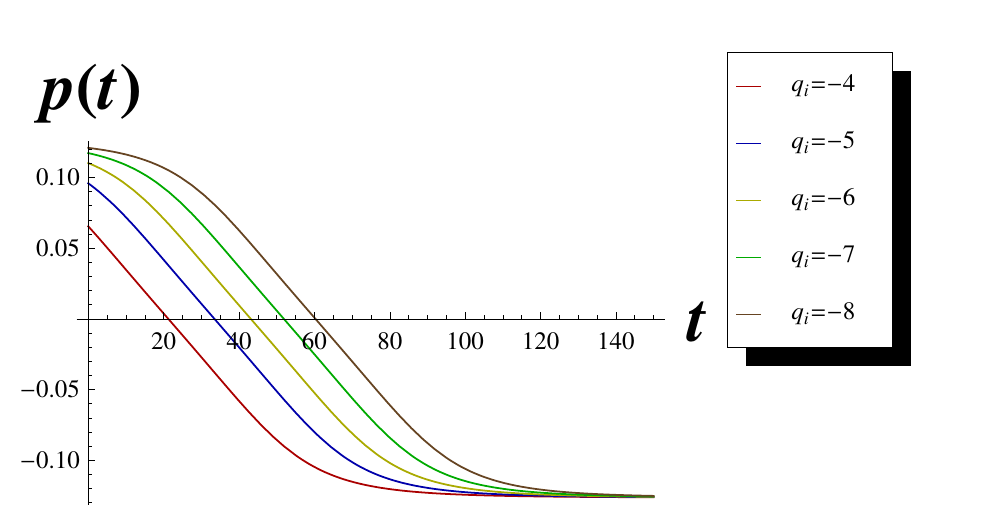}
		\caption{Classical momentum for $\gamma=1.46484$.}
		\label{class_mom_2}
	\end{figure}

Classical evolution can be obtained by numerically solving the previous equations of motion. In Figs.\ref{class_traj_1}-\ref{class_mom_2} we present trajectories for $\gamma=1$ and $\gamma > 1$, keeping the remaining parameters fixed.

\begin{multicols}{2}
\end{multicols}

\begin{multicols}{2}
\end{multicols}

The classical behavior can be observed in these trajectories: for the critical case, Fig.\ref{class_traj_1} and Fig.\ref{class_mom_1} ($\gamma=1 \iff V_0=E$) there is no barrier penetration, particles coming from the left approach asymptotically the return point as their momentum vanishes. For the classical forbidden region, Fig.\ref{class_traj_2} and Fig.\ref{class_mom_2}, ($\gamma >1 \iff V_0<E$) the reflection is complete, particles bounce back the barrier and their momentum changes sign. The initial position and the time it takes for the particle to reach the potential are dynamically irrelevant, but at the quantum level they will be essential, as we will see.

\section{Semiclassical evolution from effective momentous approach} \label{trajectories}
\subsection{Semiclassical description}
We proceed to investigate the interaction, at the quantum level, of the particles and the potential barrier by analyzing their trajectories, as we did for the classical particle in section \ref{classical}. To this end, we use the formalism presented in section  \ref{method}. The quantum corrected Hamiltonian (\ref{Hamiltonian_eff}) reads
\begin{equation}
\label{H_Q}
H_{Q}\,=\,H_{\textrm{class}}\,+\,\displaystyle\frac{1}{2\,m}\,G^{0,2}\,+\,\displaystyle\sum\limits_{n\,=\,2}^{\infty}\displaystyle\frac{1}{n!}\,\displaystyle\frac{d^{n}\,V}{d\,q^{n}}\,G^{n,0}.
\end{equation}
In this framework quantum dynamics is replaced by a Hamiltonian system with infinitely many variables, two of them the canonical pair ($q,p$), and infinitely many quantum dispersions acting as corrections in $H_Q$. For the equation of motion of $p$, Eq. (\ref{eom_p_class}), the following correction is obtained
\begin{equation}
\dot{p} = \{p,H_Q\}\,=\,-\,V'+ \sum_{k=2}^{\infty} \frac{1}{k!} G^{k,0} \left\{ p,\frac{d^k V}{d q^n} \right\} , \label{eom_p}
\end{equation}
while the equation for $q$ does not change.

To study the evolution of the system and the tunnel effect we require to solve equations (\ref{eom_q}) and (\ref{eom_p}), along with the equations for the dispersions $G^{ab}$, and suitable initial conditions. However, the complete dynamical system consists of infinitely-many differential equations which, given the form of the potential (\ref{potential}), are impossible to solve analytically so we need to consistently truncate it.

\subsection{Second order truncation}
Let us consider the system given by (\ref{Hamiltonian_eff}) and perform a truncation to second order.\footnote{The order of the truncation is considered in terms of the dispersions, that is, order $n$ for $G^{a,b}$ with $a+b \leq n$.} The effective Hamiltonian reads
\begin{equation}
\label{H_2nd}
H_{Q}\,=\frac{p^2}{2m}+ V(q)+\,\displaystyle\frac{G^{0,2}}{2\,m}\,+\,\displaystyle\frac{1}{2}\,V''\,G^{2,0}.
\end{equation}
From this equations of motion follow
\begin{eqnarray}\label{eom}
\dot{q} & = & \displaystyle\frac{p}{m}; \nonumber\\
\dot{G}^{2,0} & = & -\,\displaystyle\frac{2}{m}\,G^{1,1};\nonumber\\
\dot{p} & = & 2\,V''\,G^{1,1};\nonumber\\
\dot{G}^{1,1} & = &-\,\displaystyle\frac{G^{0,2}}{m}\,+\,V''\,G^{2,0}.
\end{eqnarray}

The effective Hamiltonian  (\ref{H_2nd}) generates a semi-classical conserved dynamical system
\begin{equation}
\begin{array}{c}
\dot{q}\dot{p}\,=\,\displaystyle\frac{d}{dt}\left(\displaystyle\frac{p^{2}}{2m}\right)\,=\,-\,\displaystyle\frac{dq}{dt}\left(\displaystyle\frac{dV}{dq}+\displaystyle\frac{1}{2}\displaystyle\frac{d}{dq}\left(V''\right)G^{2,0}\right)\,=\,-\,\displaystyle\frac{dq}{dt}\left(\displaystyle\frac{dV}{dq}+\displaystyle\frac{1}{2}\displaystyle\frac{d}{dq}(V''G^{2,0})-\displaystyle\frac{1}{2}V''\displaystyle\frac{dG^{2,0}}{dq}\right);\\
\\
\displaystyle\frac{d}{dt}\left(\displaystyle\frac{p^{2}}{2m}\right)\,=\,-\displaystyle\frac{dq}{dt}\displaystyle\frac{d}{dq}\left(V+\displaystyle\frac{G^{2,0}}{2}V''\right)+\displaystyle\frac{1}{2}V''\dot{G}^{2,0}\,=\,-\displaystyle\frac{d}{dt}\left(V+\displaystyle\frac{G^{2,0}}{2}V''\right)-\displaystyle\frac{2\,V''}{2m}G^{1,1};\\
\\
\displaystyle\frac{d}{dt}\left(\displaystyle\frac{p^{2}}{2m}\,+\,V\,+\,\displaystyle\frac{1}{2}\,V''\,G^{2,0}\,+\,\displaystyle\frac{G^{0,2}}{2\,m}\right)\,=\,\dot{H}_{Q}\,=\,0.
\end{array}
\end{equation}
It is also possible to define an effective potential\cite{Goldstein} 
\begin{equation}
\label{effective_potential}
V_{\textup{eff}}\,=\,V\,+\,\displaystyle\frac{1}{2}\,V''\,G^{2,0}\,+\,\displaystyle\frac{G^{0,2}}{2\,m}.
\end{equation}
From this expression one can give an interpretation for the behavior of the system under different conditions, in a similar way as in.\cite{Dewdney}

We can see that the quantum dispersions strongly back-react on the classical system, making the evolution nontrivial. Although we found the corresponding effective Hamiltonian and equations of motion at this order the system is not analytically solvable, and a thorough numerical exploration is required.
\subsection{Gaussian wavepacket} \label{gauss}
In order to perform a numerical evolution for the second order system (\ref{eom}) it is necessary to provide initial conditions for the dynamical system

\begin{equation}
\begin{array}{l}
q_{0}\,=\,\langle\,\hat{q}\,\rangle_{t\,=\,0};\\
\\
p_{0}\,=\,\langle\,\hat{p}\,\rangle_{t\,=\,0};\\
\\
G_{0}^{2,0}\,=\,G(0)^{2,0}\,=\,\langle\,\Delta\,\hat{q}^{2}\,\rangle_{t\,=\,0};\\
\\
G_{0}^{0,2}\,=\,G(0)^{0,2}\,=\,\langle\,\Delta\,\hat{p}^{2}\,\rangle_{t\,=\,0};\\
\\
G_{0}^{1,1}\,=\,G(0)^{1,1}\,=\,\textup{Re}\left(\left\langle \Delta\,\hat{q}\,\Psi_{0}\,|\,\Delta\,\hat{p}\,\Psi_{0} \right\rangle\right).
\end{array}
\end{equation}
We consider an initial gaussian state from which expectation values can be obtained\cite{Brizuela}. That is
\begin{equation} \label{gauss_packet}
\Psi_{0}\,=\,\displaystyle\frac{1}{(2\,\pi\,\sigma_{0}^{2})^{1/4}}\,\exp\left(-\displaystyle\frac{(q-q_{0})^{2}}{4\,\sigma_{0}^{2}}\,+\,i\,\displaystyle\frac{p_{0}}{\hbar}\,(q\,-\,q_{0})\right),
\end{equation}
$q_0$ is the peak value and $\sigma_{0}\,=\,\sqrt{G^{2,0}_{0}}$ the standard deviation. Its probability density is
\begin{equation}
\rho_{0}(q)\,=\,|\Psi_{0}|^{2}\,=\,\displaystyle\frac{1}{\sqrt{2\,\pi\,\sigma_{0}^{2}}}\,\exp\left(-\,\frac{(q-q_{0})^{2}}{2\,\sigma_{0}^{2}}\right).
\end{equation}
In this way initial conditions for the momenta $G^{0,2}$, $G^{2,0}$ and dispersion $G^{1,1}$ can be obtained from the parameters above
\begin{equation}
\label{EE1}
\begin{array}{l}
\Delta\,\hat{q}\,\Psi_{0}\,=\,(q\,-\,q_{0})\,\Psi_{0};\\
\\
\Delta\,\hat{p}\,\Psi_{0}\,=\,\displaystyle\frac{i\,\hbar}{2\,\sigma_{0}^{2}}(q-q_{0})\,\Psi_{0}\,=\,\displaystyle\frac{i\,\hbar}{2\,\sigma_{0}^{2}}\,\Delta\,\hat{q}\,\Psi_{0},
\end{array}
\end{equation}
thus
\begin{equation} \label{initial_conds}
\begin{array}{l}
G^{0,2}_{0}\,=\,\langle\,\Delta\,\hat{p}^{2}\,\rangle_{t=0}\,=\,|\,\Delta\,\hat{p}\,\Psi_{0}|^{2}\,=\,\left(\displaystyle\frac{\hbar}{2\,\sigma_{0}^{2}}\right)^{2}\,\langle\,\Delta\,\hat{q}^{2}\,\rangle\,=\,\left(\displaystyle\frac{\hbar}{2\,\sigma_{0}}\right)^{2};\\
\\
G^{1,1}_{0}\,=\,\textup{Re}\left(\langle\,\Delta\,\hat{q}\,\Psi_{0}\,|\,\Delta\,\hat{p}\,\Psi_{0}\,\rangle\right)\,=\,\textup{Re}\left(\displaystyle\frac{i\,\hbar}{2\,\sigma_{0}^{2}}\,\langle\,\Delta\,\hat{q}^{2}\,\rangle\right)\,=\,\textup{Re}\left(i\,\displaystyle\frac{\hbar}{2}\right)\,=\,0.
\end{array}
\end{equation}
To meet the above initial conditions, it is sufficient to saturate the uncertainty relation (\ref{uncertainty}), that is
\begin{equation}
G^{0,2}_{0}\,G^{2,0}_{0}\,-\,G^{1,1}_{0}\,=\,\displaystyle\frac{\hbar^{2}}{4}.
\end{equation}

\section{Numerical evolution} \label{numerics}

We are in position to study the evolution of the dynamical system governed by equations (\ref{eom}). For generic potential barriers of the form (\ref{potential}) it is not possible to obtain analytical solutions, but we can perform a numerical evolution and study its behavior. We determine individual semiclassical trajectories, in a similar fashion as those discussed in the Bohmian quantum mechanics\cite{Bohm}.

Given the initial conditions (\ref{initial_conds}) we generate trajectories for particles that come from the left with energy $E_0$. We also need initial conditions for the classical configuration variables ($q,p$), discussed below. Evolution is restricted to physical trajectories, that is, those constrained by the uncertainty relation (\ref{uncertainty}).

Tunneling occurs when a particle reaches points beyond the classical return points $x$ (\ref{return_points}), and we can distinguish two types of tunneling: one where the particle goes beyond the turning point $x>0$ on the right and can be detected, and a second where the particle moves to the right of the left turning point but gets ``trapped" inside the potential, $|q|< x$. Such behavior has been previously discussed (see for instance\cite{Dewdney}). Initial conditions for quantum dispersions are obtained by preparing gaussian packets with parameters ($q_0, p_0, \sigma_0$). 
\subsection{Semiclassical trajectories}
\begin{figure}[H] 
	\centering
	\includegraphics[scale=0.65]{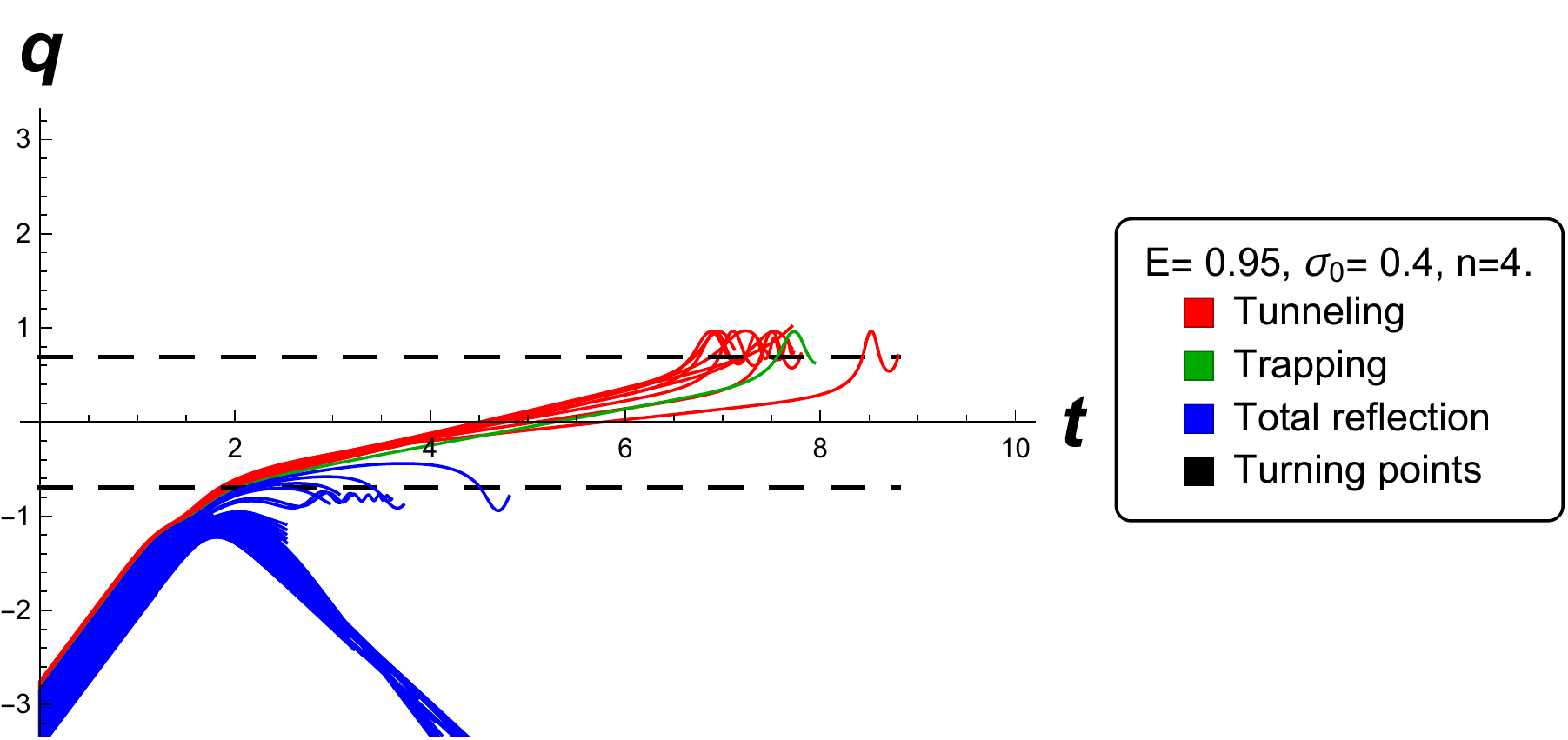}
	\caption{Trajectories for $-3.72638<q_{0}<-1.36634$. Dashed lines indicate classical return points.}
	\label{semiclass_traj_1}
\end{figure}

For generic initial conditions, such as those presented above, we obtain three different behaviors (Fig.\ref{semiclass_traj_1})

{\bf Reflection}. Blue curves show the evolution of particles that are fully reflected, as in the classical case. Particles whose energy lies below the classical potential, $\gamma>1$, are reflected at points to the left of the classical return point. This is because they interact with the effective time-dependent potential.

{\bf Tunneling}. A second distinctive behavior is when particles interact with the potential barrier and go through to the right, displaying its true quantum behavior and effectively tunneling. Trajectories of this kind are shown in red. The interaction with the effective potential seemingly occurs before the classical left turning point, as in the previous case.

\begin{figure}[H]
	\centering
	\includegraphics[scale=0.6]{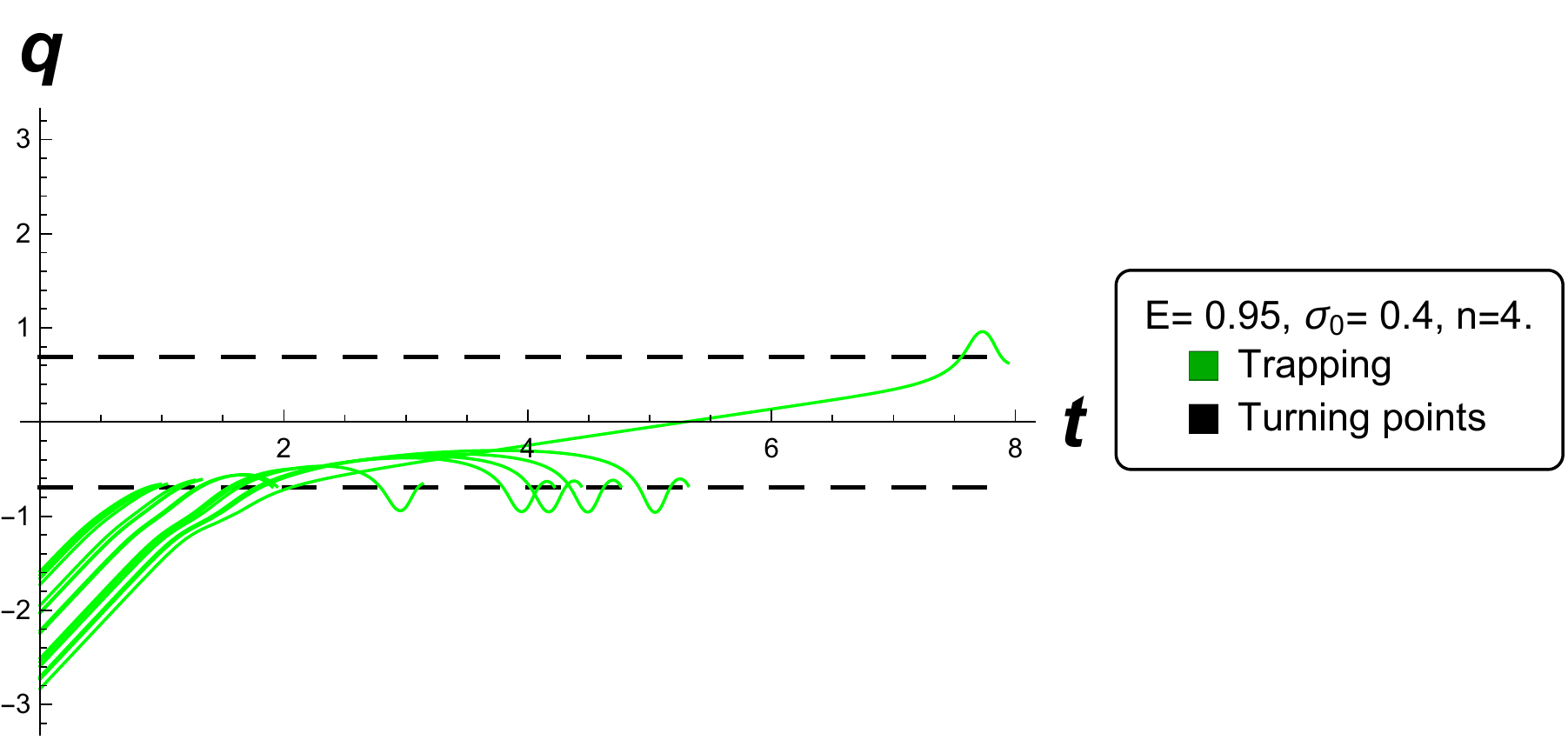}
	\caption{Particles trapped near the turning points.}
	\label{trap_1}
\end{figure}

\begin{figure}[H]
	\centering
	\includegraphics[scale=0.6]{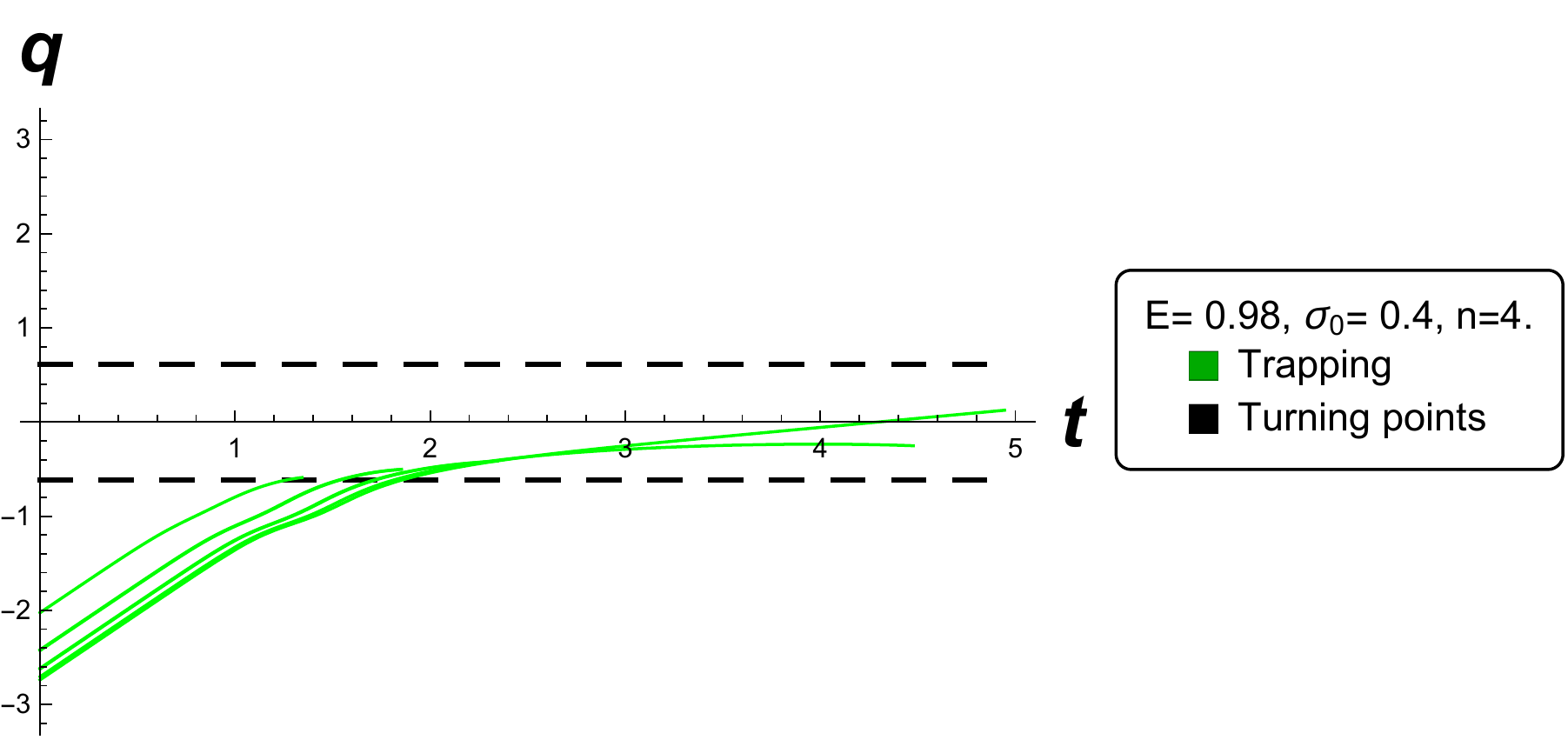}
	\caption{Particles trapped deep into the potential.}
	\label{trap_2}
\end{figure}

\begin{figure}[H]
	\centering
	\includegraphics[scale=0.6]{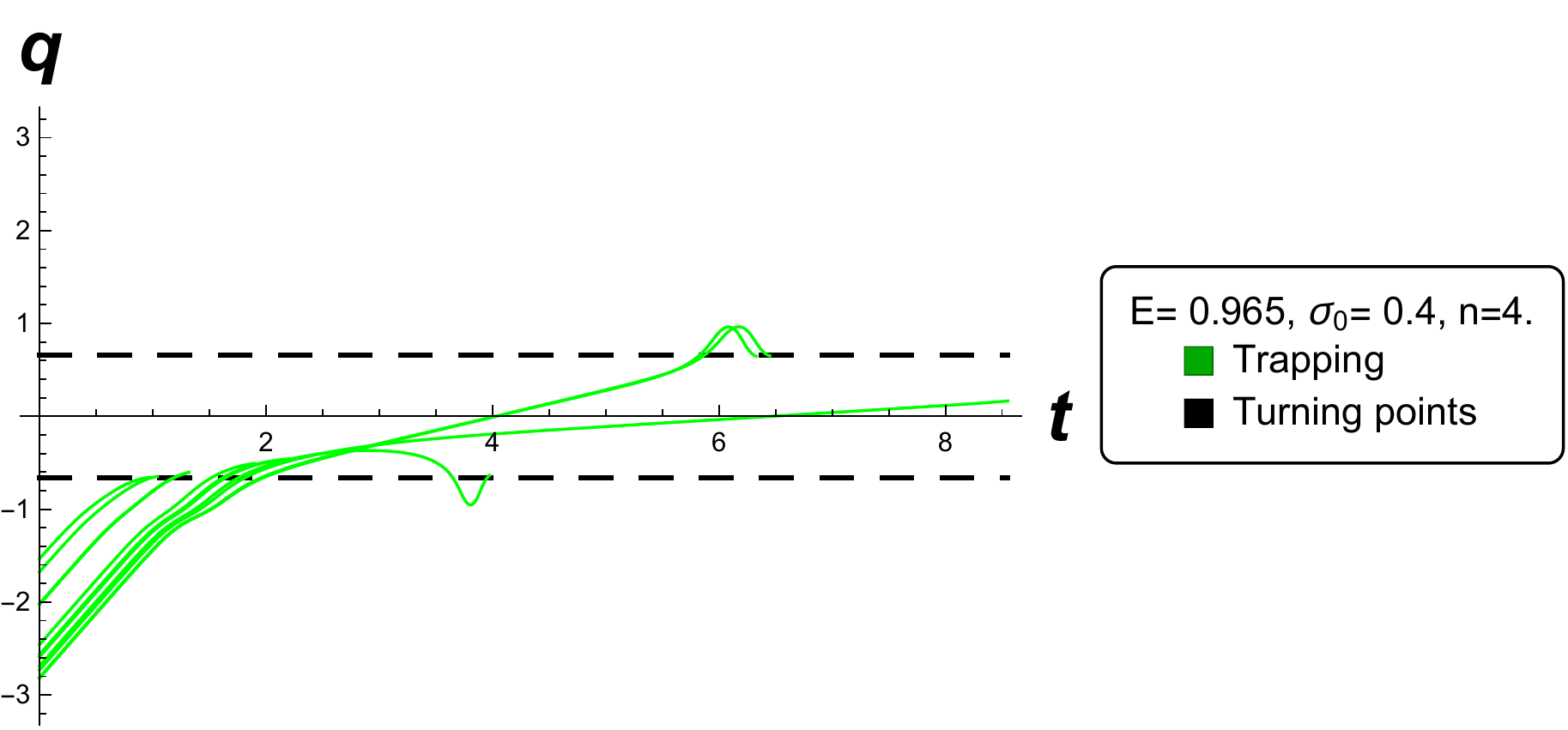}
	\caption{Semiclassical trapping of particles inside the effective quantum potential.}
	\label{trap_3}
\end{figure}

\begin{figure}[H]
	\centering
	\includegraphics[scale=0.6]{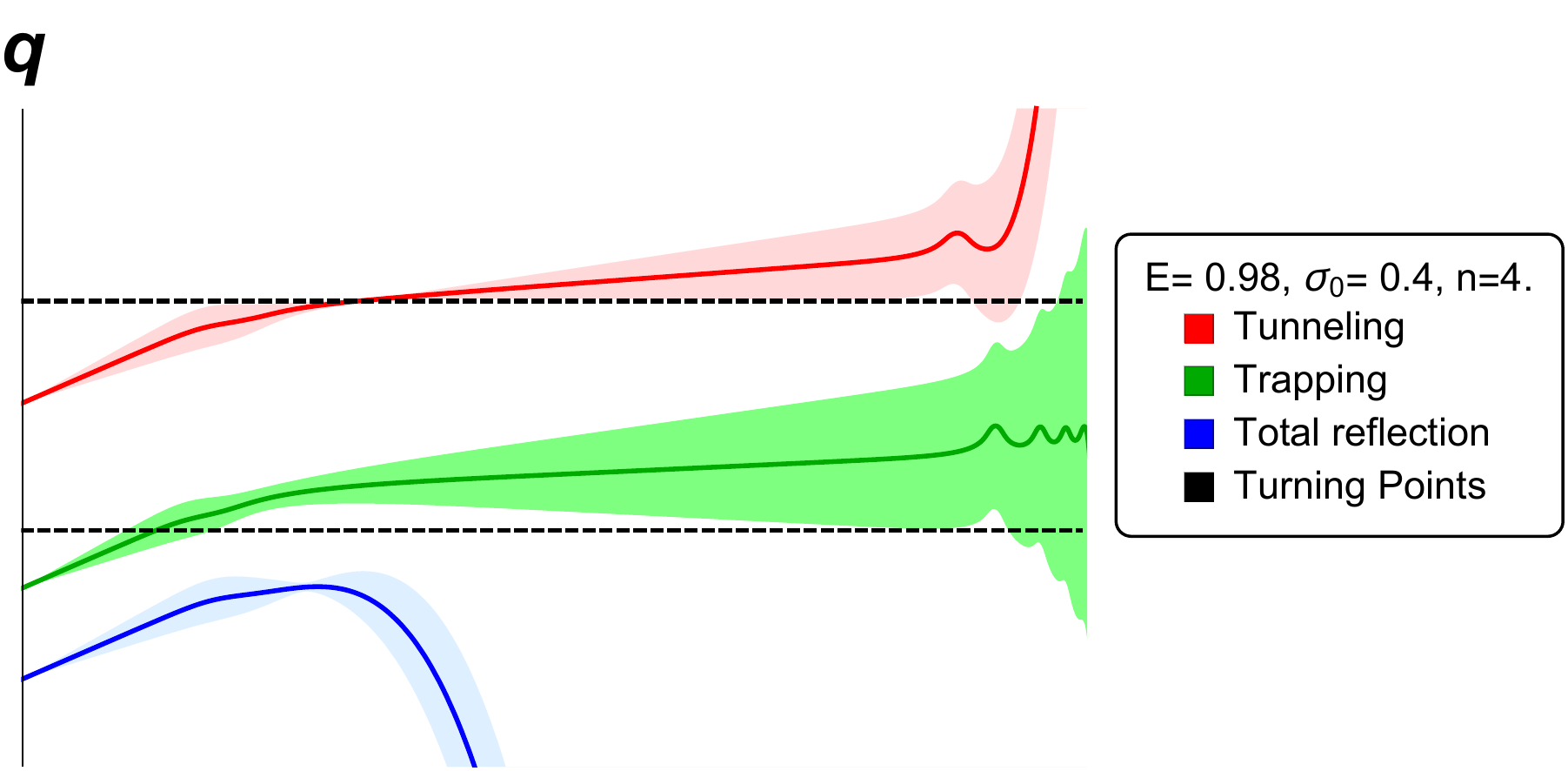}
	\caption{Semiclassical trajectories bounded by their quantum dispersions.}
	\label{shadow_turning_points}
\end{figure}

{\bf Trapping}. A third behavior is when the particles get ``trapped", shown in green. Classically these particles would reflect or remain at the turning point in an unstable equilibrium for $\gamma=1$, but at the quantum level they get trapped within the barrier (Figs.\ref{trap_1}-\ref{trap_3}). Their trajectories bounce back and forth inside the potential barrier. All this can be explained in terms of the effective potential.
\vspace{-1.5cm}
In Fig. \ref{shadow_turning_points} we show the semiclassical trajectories bounded by its dispersion, shown as a shadow, $q(t)\pm \sqrt{G^{2,0}(t)}$. From (\ref{eom}) it is clear that the quantum dispersions are dynamical, as can be seen in the figure, so the validity of the numerical evolution must be constrained by (\ref{uncertainty}). 

\subsection{Effective potential dynamics}
The results obtained in this section can be interpreted by analyzing the dynamics generated by the effective potential (\ref{effective_potential}). We consider the case $n=4$ since it is particularly illuminating and it contains all the features of higher order cases.

\begin{figure}[H]
	\centering
	\includegraphics[scale=0.9]{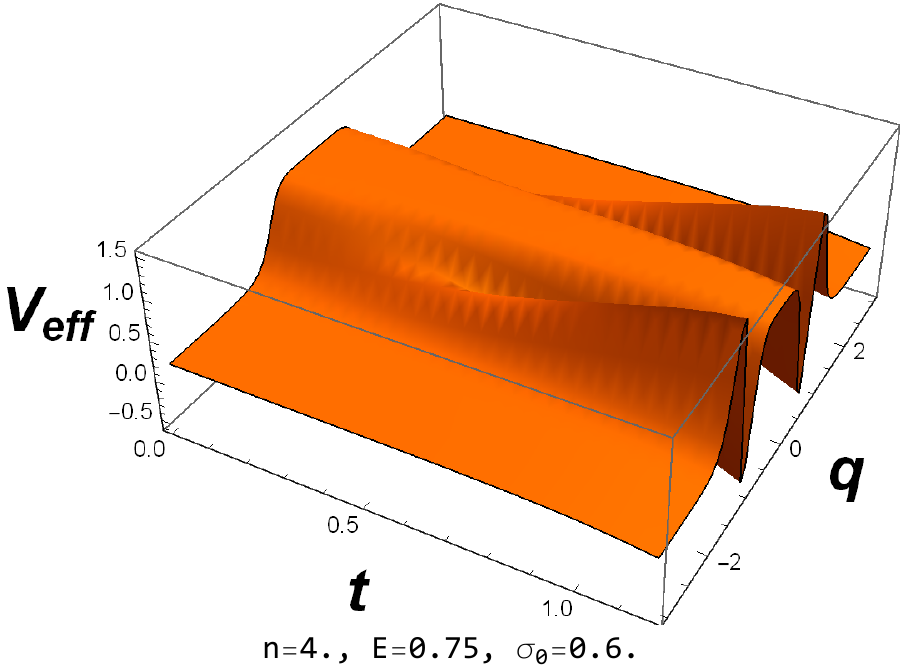}
	\caption{Effective potential $V_{\textup{eff}}(q,t)$ as a function of time and position.}
	\label{potential_3D}
\end{figure}

\begin{figure}[H] 
	\centering
	\includegraphics[scale=0.7]{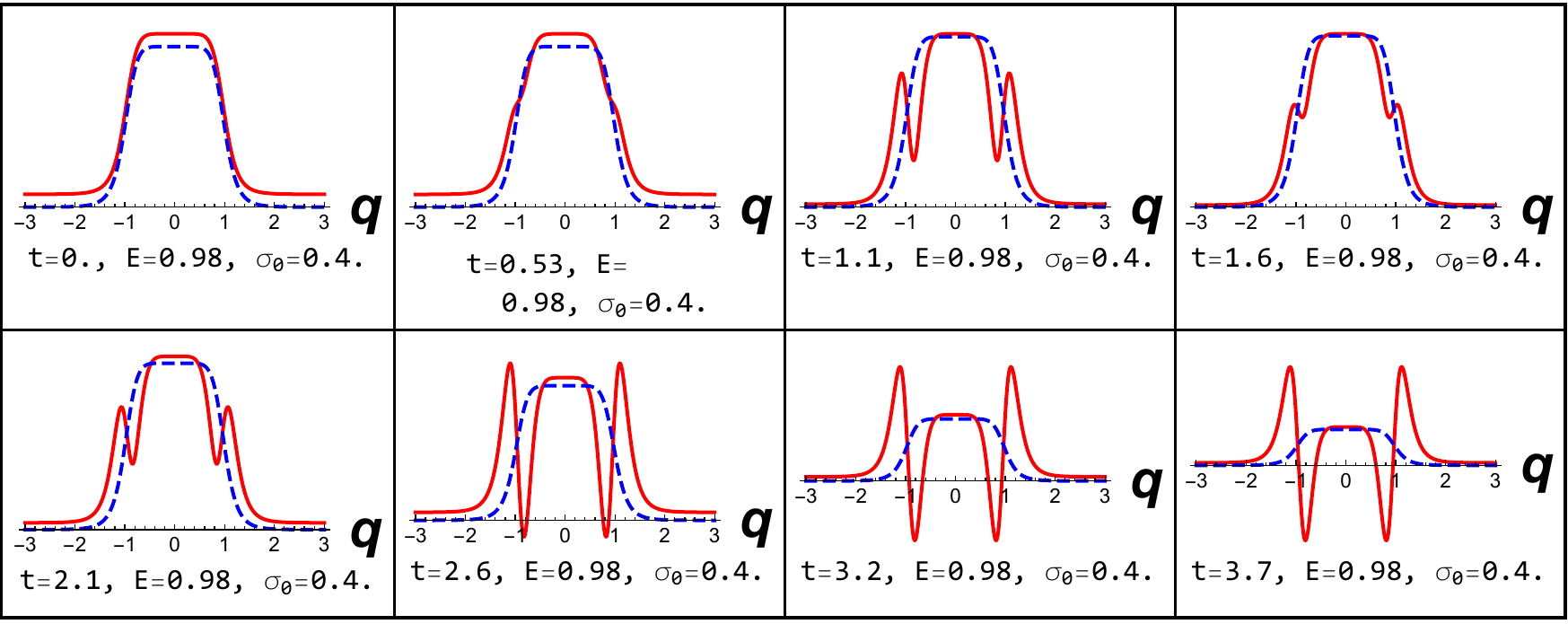},
	\caption{ Constant time sections of $V_{\textup{eff}}$ (red) for $t\,=\,0$ (first) up to $t\,=\,3.7$ (last), for $\alpha=1$ and $E=0.98$ from left to right, $V_{\textrm{class}}$  is shown in dashed blue.}
	\label{V_eff_2D}
\end{figure}

In Fig. \ref{potential_3D} we show a three dimensional plot for the effective time dependent potential $V_{\textrm{eff}}(q,t)$.
In Fig.\ref{V_eff_2D} we show the classical potential in dashed blue and some sections of $V_{\textrm{eff}}(q,t)$ at several different times. 

The interpretation of the previous semiclassical evolution is as follows: the classical potential (\ref{potential}) is modified due to the quantum back-reaction of quantum dispersions yielding the effective potential $V_{\textrm{eff}}(q,t)$,  which in turn evolves and moves away from the former. Initially, $V_{\textrm{eff}} > V_{\textrm{class}}$, then it decreases for a short period of time, then it bounces back up in a symmetrical way and, for later times, it acquires two new regions of minimal potential where stability can be attained. Incident particles from the left find the potential at different heights depending on their initial conditions. 

There are three different possibilities
\begin{enumerate}
	\item Potential higher than the incident energy: the particle is reflected;
	\item Potential lower than the incident energy: if the particle travels through the potential while it is lower than its energy, then the particle can go through the potential and emerge on the right, effectively tunneling;
	\item Potential lower than the incident energy: the particle penetrates into the potential, but as the later increases in height the particle gets trapped in one of the two potential minima. 
\end{enumerate}

Indeed, if the particle penetrates the barrier and the time it takes to go through it is shorter than the time required for the barrier to grow higher than the particle's energy there may be transmission. This also explains the existence of a trap between the two potential valleys where the particle bounces back and forth, the particle was able to penetrate the barrier from the left but the potential grew fast enough that the particle could not go over it.

\begin{figure}[H]
		\centering
		\includegraphics[scale=0.5]{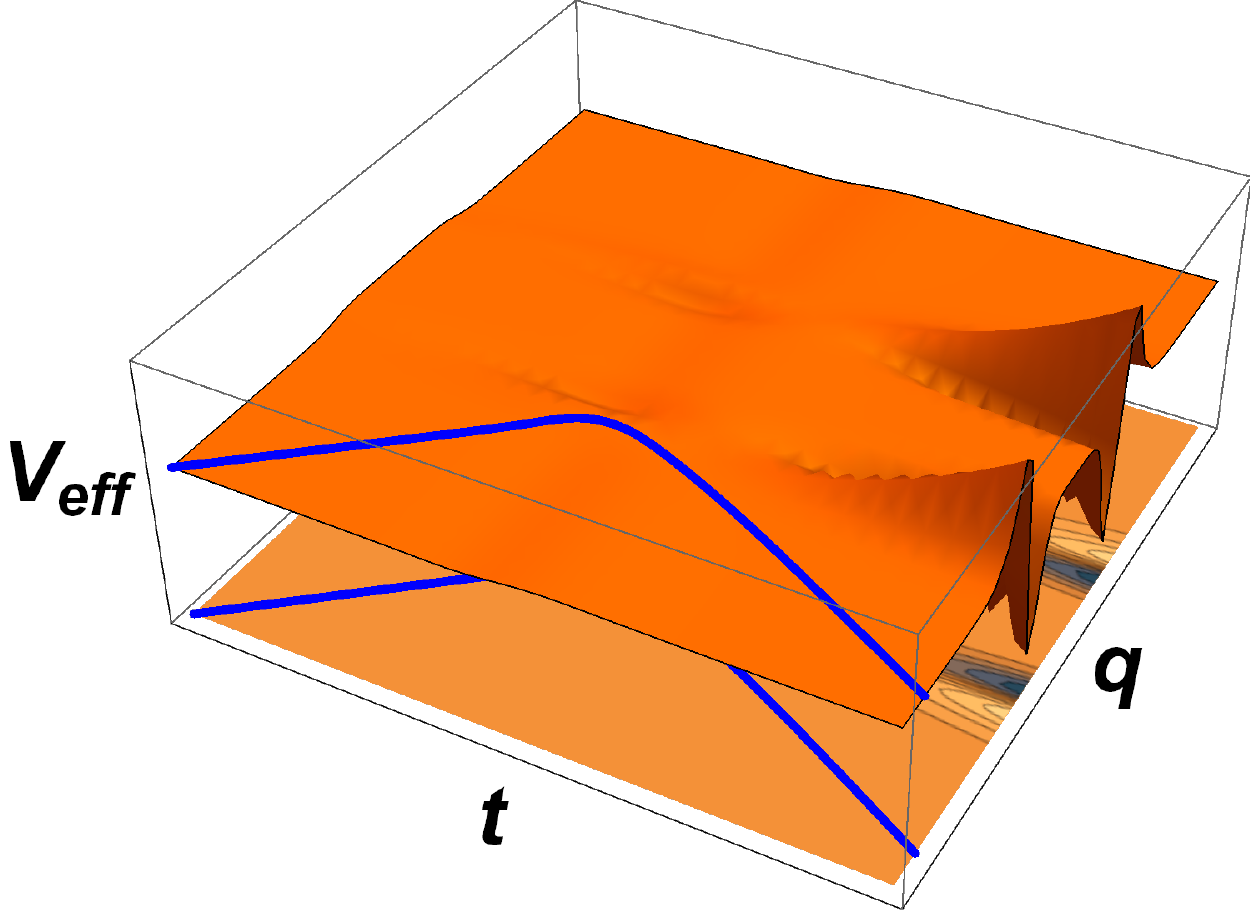}
		\caption{Reflection}
		\label{reflection_3d}
	\end{figure}

\begin{figure}[H]
		\centering
		\includegraphics[scale=0.5]{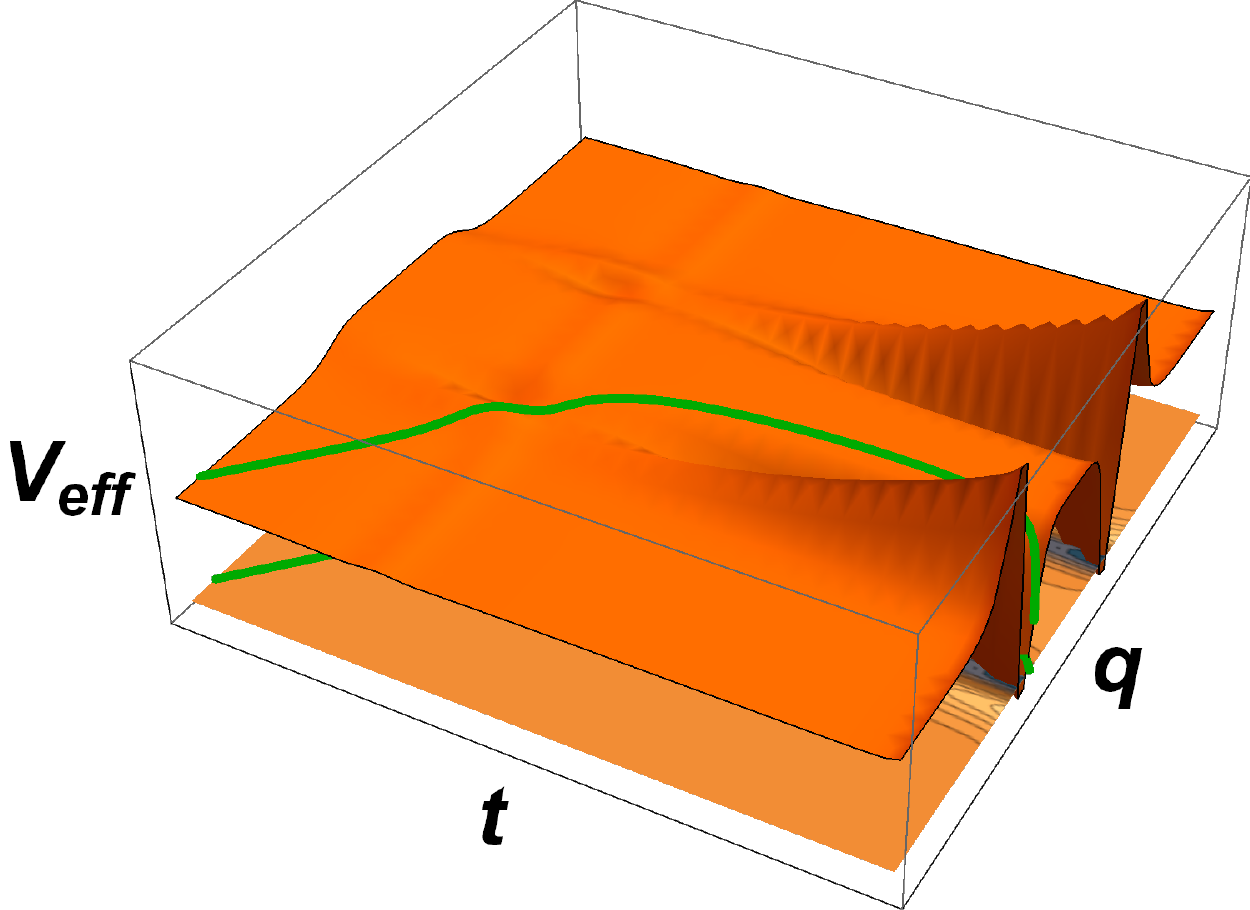}
		\caption{Trapping}
		\label{trap_3d}
	\end{figure}

\begin{figure}[H]
		\centering
		\includegraphics[scale=0.5]{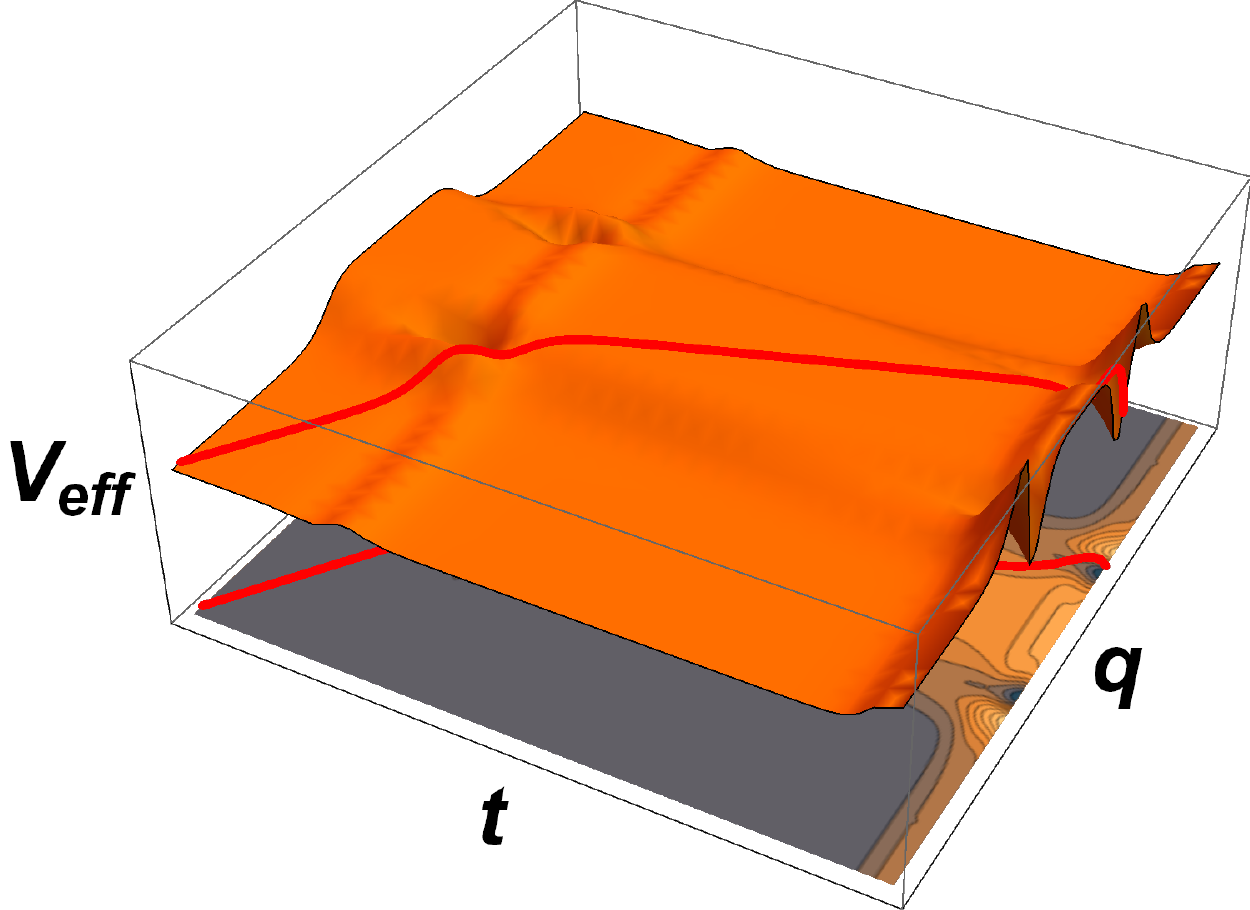}
		\caption{Tunneling}
		\label{tunnel_3d}
\end{figure}

\begin{figure}[H]
		\centering
		\includegraphics[scale=0.4]{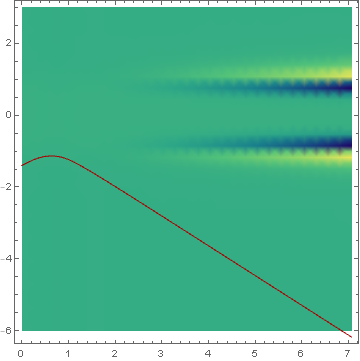}
		\caption{Reflection near $V_{\textup{eff}}(0)$}
		\label{density_reflection}
	\end{figure}

\begin{figure}[H]
		\centering
		\includegraphics[scale=0.4]{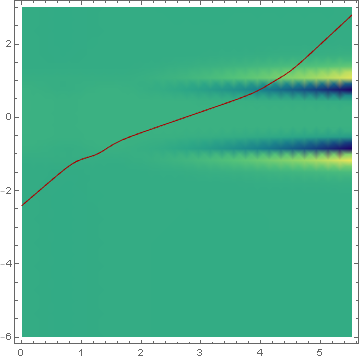}
		\caption{Tunneling near $V_{\textup{eff}}(0)$.}
		\label{density_tunnel}
	\end{figure}

\begin{figure}[H]
		\centering
		\includegraphics[scale=0.4]{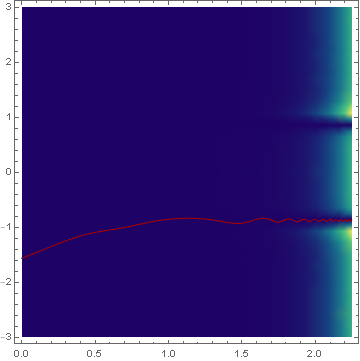}
		\caption{Trapping inside $V_{\textup{eff}}$, left valley.}
		\label{density_trapp_1}
	\end{figure}

\begin{figure}[H]
		\centering
		\includegraphics[scale=0.4]{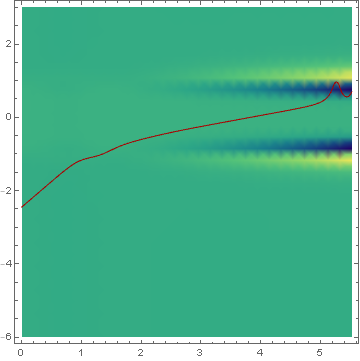}
		\caption{Trapping inside $V_{\textup{eff}}$, right valley.}
		\label{density_trapp_2}
	\end{figure}

We illustrate this in the following 3D plots, Figs.\ref{reflection_3d}-\ref{tunnel_3d}, where reflection, trapping and tunneling for different initial conditions is displayed.
\begin{multicols}{2}
\end{multicols}
We also show the different trajectories in density plots. In Fig. \ref{density_reflection} a particle incident from the left gets reflected by $V_{\textrm{eff}}(q,t)$ as the potential is higher than its energy, whereas in Fig. \ref{density_tunnel} the particles tunnels through the effective potential as its height is lower than $V_{\textrm{eff}}(0)$.
\begin{multicols}{2}
\end{multicols}
Fig. \ref{density_trapp_1}  and Fig. \ref{density_trapp_2} exhibit the tunneling and trapping of particles inside the two valleys of $V_{\textrm{eff}}$. As can be seen, when the particles get trapped inside the potential they behave as classical dynamics dictates, they oscillate around the potential minimum.
\pagebreak

\begin{multicols}{2}
\end{multicols}

The final fate of the particle is completely probabilistic. Due to the uncertainty in the position, it is not possible to know exactly how it behaves, but it will be reflected from the barrier if it was trapped in the left valley, while it will tunnel away if it was trapped in the right valley.

\section{Discussion} \label{discussion}

The physical tunneling process, considered a genuine quantum mechanical effect, is usually studied in quantum mechanics employing time-independent or semi-classical methods such as WKB, or numerically by computing the transmission-reflection coefficients for the corresponding system. In this work we studied the tunnel effect within a detailed semi-classical approach, the so-called momentous quantum mechanics, where it is possible to obtain effective individual trajectories for the particles providing a more intuitive description of the process. These ``effective trajectories" resemble those appearing in the Bohm's formulation of quantum mechanics\cite{Prezhdo,Sanz1,Sanz2}.

With this approach we were able to generate numerically particle trajectories for different initial conditions, as well as the dispersions that back-react on the classical evolution. In this way it is possible to identify which particles will cross or reflect from the barrier. It is also possible to track back the preparation of the system or the experiment, measuring or detecting the transmission fraction, thus providing an interesting experimental tool. In the present formulation, the definition of transmission and reflection coefficients is very natural, simply counting the transmitted or reflected trajectories. However we do not obtain a completely deterministic evolution for the particles, their behavior is subject to quantum uncertainty because the dispersion variables defining the quantum state of the system are dynamic and constrain the evolution.

A remarkable feature of this effective formulation is that one is able to determine three different possibilities for the evolution of a particle, it can cross the barrier, get reflected by it, or get trapped inside the potential. This last possibility exists because the classical potential barrier is modified, decreased by the back-reaction of quantum dispersions, generating an effective quantum potential in resemblance to the potential obtained in the Bohm's formulation\cite{Dewdney}. Initially, the effective potential decreases from its initial value and then changes such that two minima appear on either side of the maximum, where a particle can live inside the potential for a short period of time. Due to quantum fluctuations the particle eventually tunnels away or gets reflected. Certainly this effective potential is interesting for it may have implications in several fields\cite{quantum_potential}. For instance it is possible to compute the time that a particle need to cross the barrier, the so-called tunneling time\cite{time-tunnel}, although a deeper analysis must be done on this matter due to the existence of uncertainty in the evolution. A discussion on the subject has been held in a recent work for a type of ionization experiment\cite{Bojo-Crowe}.

In this sense our analysis may be an interesting starting point for research at the applied level. Even though quantum trajectories are neither observable nor experimentally measurable, the information provided by the present formulation can be used as a tool for studying experimental processes whenever a quantum tunneling is present, as is the case in quantum control experiments\cite{quantum_control}, multielectron and ionization experiments\cite{ionization}, among others. Indeed the trapping feature of the effective potential may have implications for resonant tunneling diodes where electron capture between potential wells is an important ingredient\cite{resonant_diods}. Our description is not complete since it is not possible to solve a system of infinite equations, however, we consider that this could be an interesting ground for testing certain hypotheses in the corresponding regimes.


\appendix{Third order evolution}
We show the evolution of the system with effective Hamiltonian (\ref{H_Q}) at third order in quantum dispersions $G^{a,b}$. The Hamiltonian reads
\begin{equation}
H_{Q} = H_{\textrm{class}}\,+\,\displaystyle\frac{G^{0,2}}{2\,m}\,+\,\displaystyle\frac{1}{2}\,V''\,G^{2,0}\,+\,\displaystyle\frac{1}{6}\,V'''\,G^{3,0}.
\end{equation}
The system consists now of classical configuration variables $(p,q)$, second order momenta $(G^{2,0}, G^{1,1}, G^{0,2})$ and third order momenta $(G^{3,0}, G^{2,1}, G^{1,2}, G^{0,3})$. The equations of motion are
\begin{equation} \label{eom_3rd}
\begin{array}{l}
\dot{q}\,=\,p/m;
\\
\dot{p}\,=\,-V'\,-\,\displaystyle\frac{1}{2}\,V'''\,G^{2,0}\,-\,\displaystyle\frac{1}{6}\,V''''\,G^{3,0};
\\
\dot{G}^{0,2}\,=\,2\,V''\,G^{1,1}\,+\,V'''\,G^{2,1};
\\
\dot{G}^{2,0}\,=\,-\,\displaystyle\frac{2}{m}\,G^{1,1};
\\
\dot{G}^{1,1}\,=\,-\,\displaystyle\frac{G^{0,2}}{m}\,+\,V''\,G^{2,0}\,+\,\displaystyle\frac{1}{2}\,V'''\,G^{3,0};
\\
\dot{G}^{1,2}\,=\,-\displaystyle\frac{G^{0,3}}{m}\,+\,2\,V''\,G^{2,1}; 
\\
\dot{G}^{2,1}\,=\,-\displaystyle\frac{2}{m}\,G^{1,2}\,+\,V''\,G^{3,0}; 
\\
\dot{G}^{3,0}\,=\,-\displaystyle\frac{3}{m}\,G^{2,1};
\\
\dot{G}^{0,3}\,=\,3\,V''\,G^{1,2}.
\end{array}
\end{equation}

Initial conditions are obtained from the gaussian wavepacket as in section \ref{gauss}. Conditions (\ref{initial_conds}) get complemented by the following
\begin{equation}
\begin{array}{l}
G^{3,0}_{0}\,=\,\langle\,\Delta\,\hat{q}^{3}\,\rangle\,=\,0;\\
\\
G^{0,3}_{0}\,=\,\langle\,\Delta\,\hat{p}^{3}\,\rangle\,=\,\displaystyle\frac{i\,\hbar}{2\,\sigma_{0}^{2}}\,\langle\,\Delta\,\hat{p}^{2}\,\Delta\,\hat{q}\,\Psi_{0}\,\rangle\,=\,-\,\displaystyle\frac{\hbar^{2}}{\sigma_{0}^{2}}\,p_{0}\,+\,\displaystyle\frac{i\,\hbar}{2\,\sigma_{0}^{2}}\,p_{0}^{2}\,\langle\,\Delta\,\hat{q}\,\rangle\,=\,-\,\displaystyle\frac{\hbar^{2}}{\sigma_{0}^{2}}\,p_{0}\\
\\
G^{1,2}_{0}\,=\,\textup{Re}\left(\langle\,\Delta\,\hat{q}\,\Psi_{0}\,|\,\Delta\,\hat{p}^{2}\,\Psi_{0}\right)\,=\,\textup{Re}\left(\langle\,\Delta\,\hat{p}^{2}\,\Delta\,\hat{q}\,\Psi_{0}\,|\,\Psi_{0}\,\rangle\right)\,=\,\textup{Re}(-i\,2\,\hbar\,p_{0})\,=\,0;\\
\\
G^{2,1}_{0}\,=\,\textup{Re}\left(\langle\,\Delta\,\hat{p}^{2}\,\Psi\,|\,\Delta\,\hat{q}\,\Psi_{0}\,\rangle\right)\,=\,\textup{Re}\left(\langle\,\Psi\,|\,\Delta\,\hat{p}^{2}\,\Delta\,\hat{q}\,\Psi_{0}\,\rangle\right)\,=\,\textup{Re}\left(i\,2\,\hbar\,p_{0}\right)\,=\,0.
\end{array}
\end{equation}

Below we show the effective potential and trajectories for the three cases: reflection, tunneling and trapping, and trajectories with uncertainty for transmission and trapping. Notice that, at this order, the behavior is quite similar to that at second order.

\begin{figure}[H]
	\centering
	\includegraphics[scale=0.65]{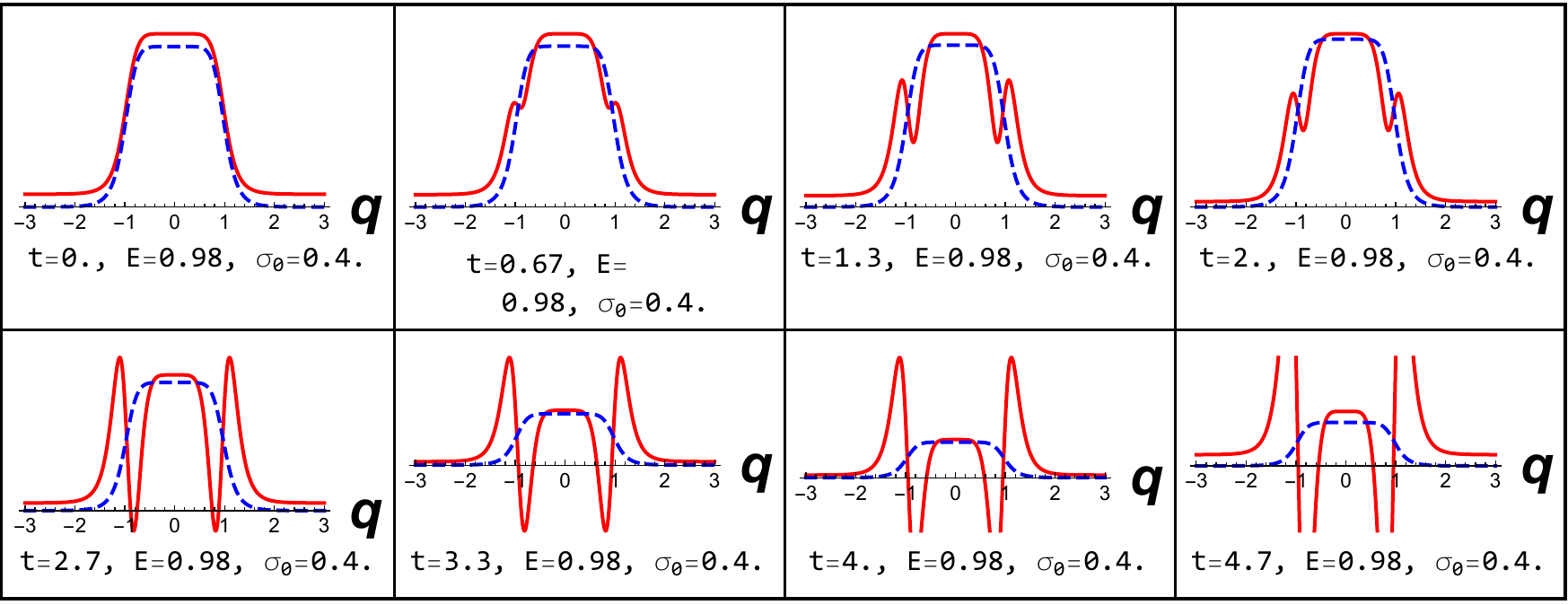}
	\caption{$V_{\textup{eff}}(q)$ in solid red, and $V_{\textrm{class}}$ in dashed blue, for some values of $t$.}
	\label{potential_3_order}
\end{figure}

\begin{figure}[H]
	\centering
	\includegraphics[scale=0.6]{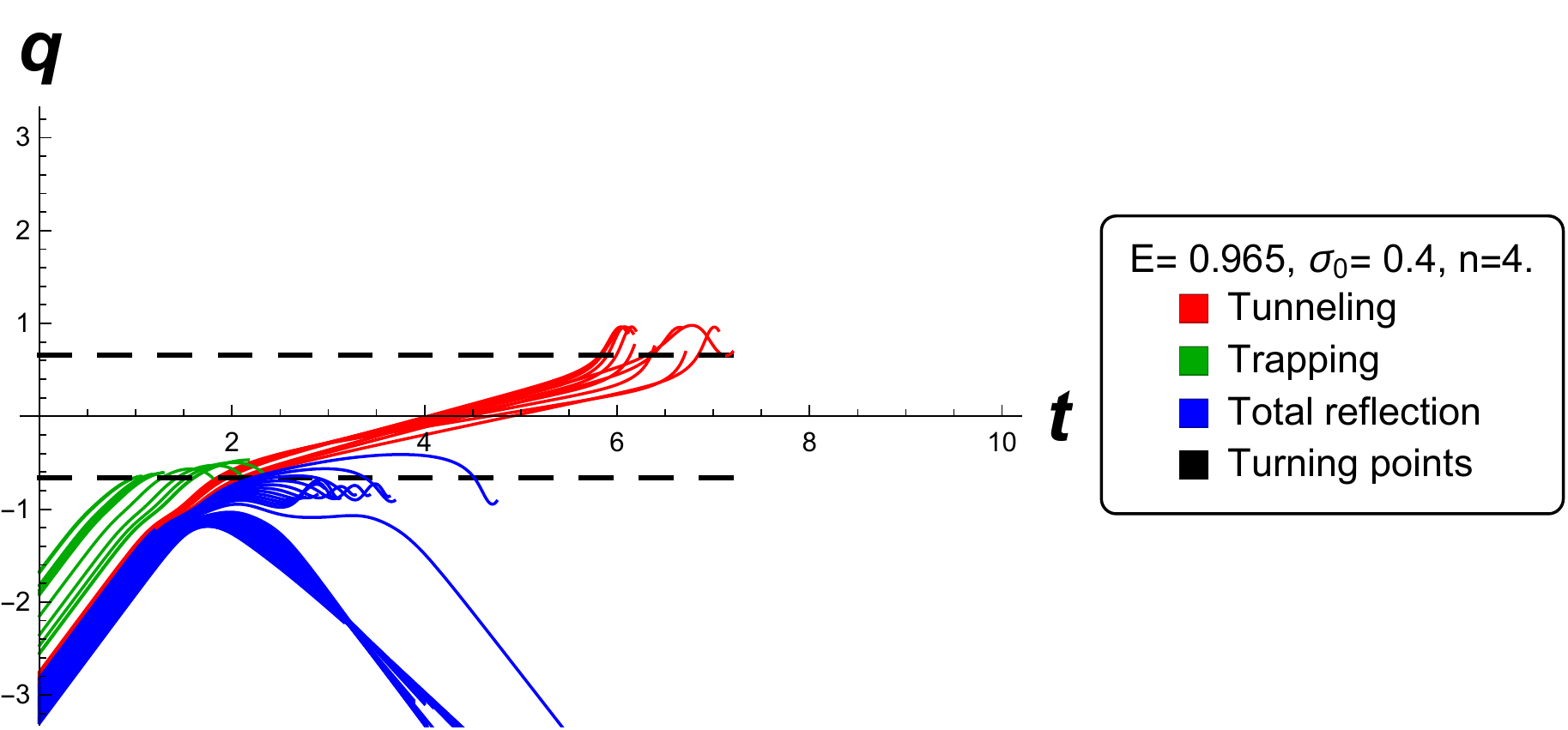}
	\caption{Semiclassical trajectories for generic conditions, displaying reflection, transmission and trapping.}
	\label{semiclass_traj_3_3order}
\end{figure}

\begin{figure}[H]
		\centering
		\includegraphics[scale=0.5]{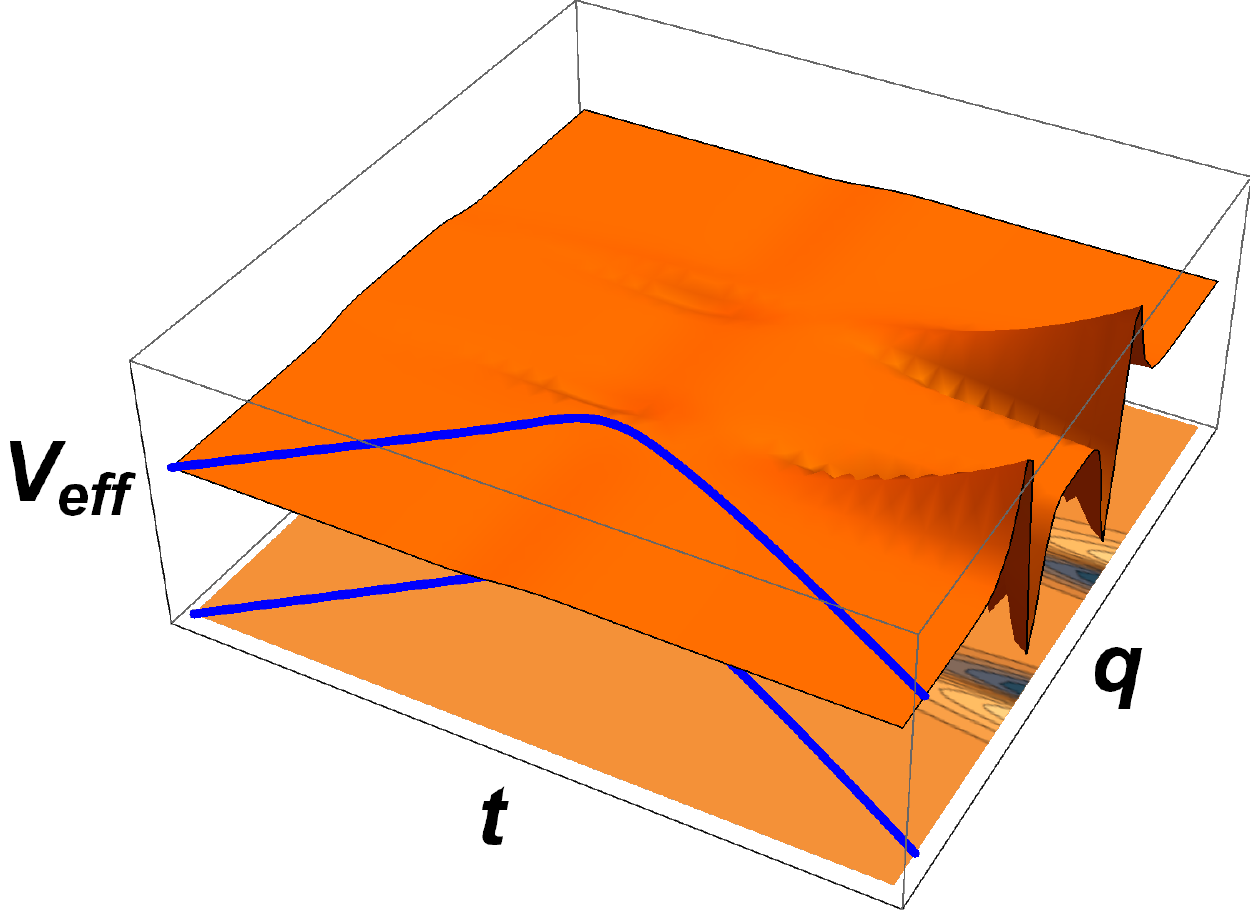}
		\caption{Reflection}
		\label{reflection_3d_3order}
	\end{figure}

\begin{figure}[H]
		\centering
		\includegraphics[scale=0.5]{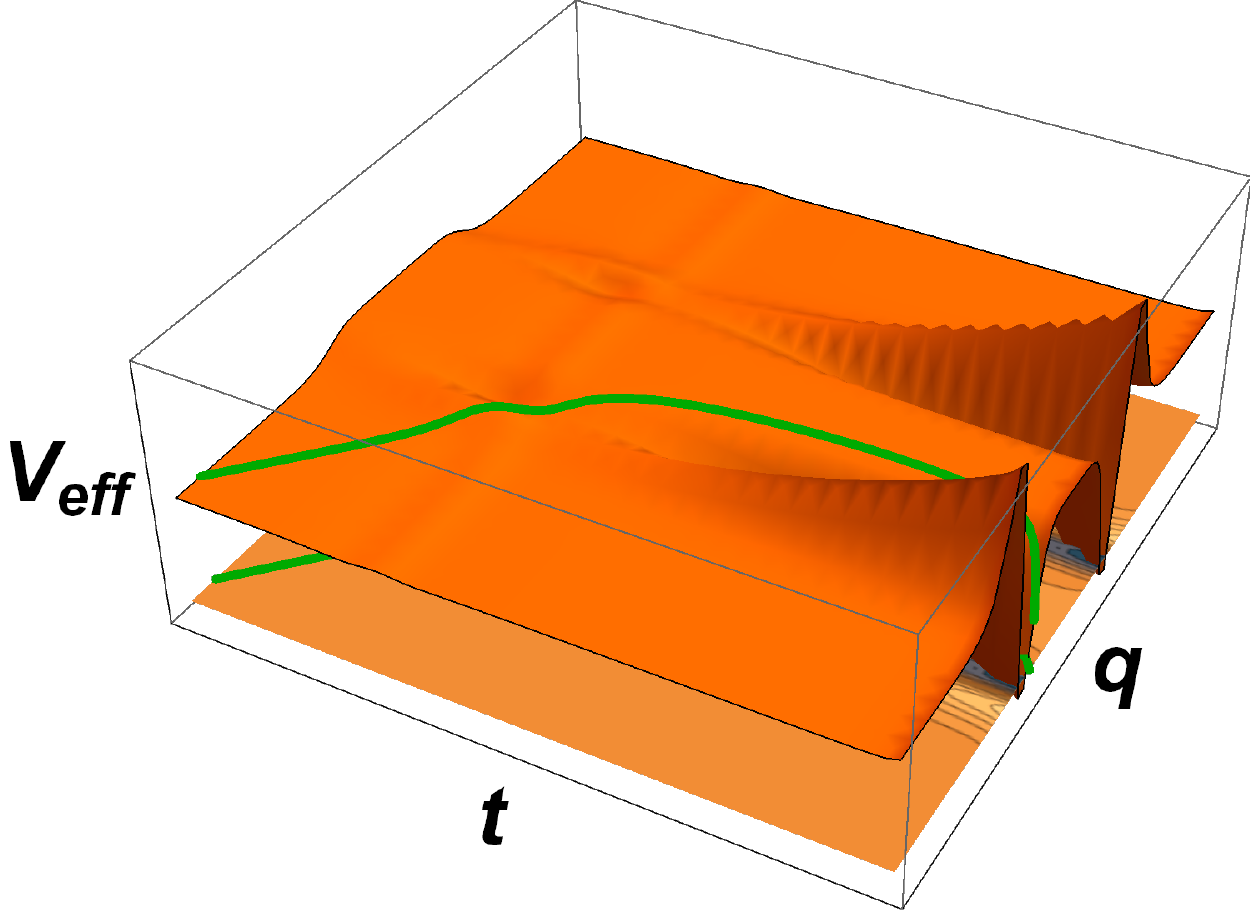}
		\caption{Trapping}
		\label{trap_3d_3order}
	\end{figure}

\begin{figure}[H]
		\centering
		\includegraphics[scale=0.5]{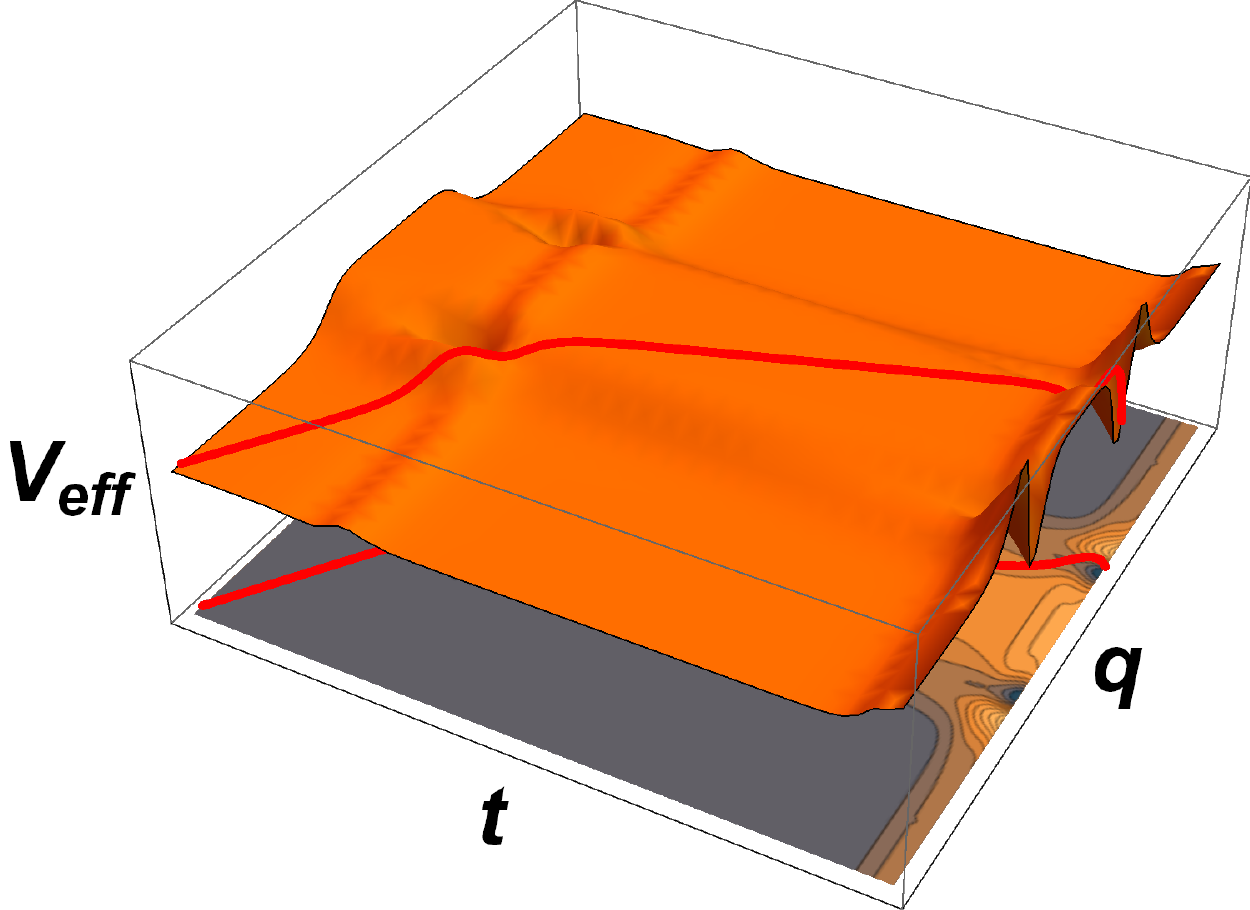}
		\caption{Tunneling}
		\label{tunnel_3d_3order}
\end{figure}

Sections for the effective potential are displayed in Fig. \ref{potential_3_order}
Trajectories for the three different behaviors are shown in Fig. \ref{semiclass_traj_3_3order}.
Three dimensional trajectories and the effective potential are shown in Figs. \ref{reflection_3d_3order} to \ref{tunnel_3d_3order}
\begin{multicols}{2}
\end{multicols}

We obtain the same general behavior as in the second order case, under the same set of initial parameters. The absolute error between the third and second order truncations is around 5 \%. Therefore to analyze the tunnel effect within the effective momentous method presented here one can stick to the second order formulation and obtain results with with high accuracy.


\end{document}